\documentclass[preprint,aps,pra,tightenlines,showpacs,endfloats,amsmath]{revtex4}
\usepackage[dvips]{graphicx}  
\usepackage{dcolumn}          
\usepackage{bm}               
\begin{document}
\def\bea{\begin{eqnarray}}
\def\eea{\end{eqnarray}}
\def\be{\begin{equation}}
\def\ee{\end{equation}}
\def\rra{\right\rangle}
\def\lla{\left\langle}
\def\ra{\rightarrow}
\def\kvp{\bm{k}'}
\def\kv{\bm{k}}
\def\qv{\bm{q}}
\def\pv{\bm{P}}
\def\hv{\bm{h}}
\def\zv{\bm{0}}
\def\de{\Delta}  
\def\om{\omega}
\def\tb{T_{BB}}
\def\tf{T_{BF}}
\def\lz{l_z} 
\def\Lix{$^{6}$Li}
\def\Lin{$^{7}$Li}
\def\Rb{$^{87}$Rb}
\def\K{$^{40}$K}

\title{Pairing in two-dimensional boson-fermion mixtures}

\author{J. Mur-Petit, A. Polls }  
\affiliation{Departament d'Estructura i Constituents de la Mat\`eria,
             Universitat de Barcelona,
             Av.~Diagonal 647, E-08028 Barcelona, Spain}

\author{M. Baldo, H.-J. Schulze}
\affiliation{Sezione INFN, Dipartimento di Fisica, Universit\`a di Catania,
             Via Santa Sofia 64, I-95123 Catania, Italy}


\begin{abstract}
The possibilities of pairing in two-dimensional boson-fermion mixtures are
carefully analyzed. 
It is shown that the boson-induced attraction between
two identical fermions dominates the $p$-wave pairing at low density. 
For a given fermion density, the pairing gap becomes maximal at a certain 
optimal boson concentration. 
The conditions for observing pairing in current experiments are discussed.
\end{abstract}

\pacs{
 03.75.Ss    
 03.75.Nt    
 74.20.Fg    
     }

\maketitle


Since the first realization of a degenerate Fermi gas \cite{demarco99}, the
search for a BCS-like transition signature in ultracold trapped gases of
fermionic atoms has received a lot of attention both from theoretical and
experimental points of view. 
Already before the experimental achievement of \cite{demarco99}, there had 
been suggestions on the possibility to observe
this transition in a gas with two hyperfine components of \Lix~\cite{hou}.
The importance of the asymmetry in the populations of the two components was
studied in \cite{russ,old}. 
Later on, the influence of adding bosons and the presence of a BEC
on the transition temperature of the Fermi gas was studied for a
three-dimensional trap in \cite{hei00,viv00,viv}.

At the same time, the possibility to design the trap so as to produce
effectively one- and two-dimensional systems has attracted much interest in
theoretically describing \cite{petrov,baran,teo} and experimentally
obtaining \cite{exp} such low-dimensional quantum systems, where correlations
play generally a more important role than in their three-dimensional
counterparts. 


In this article we discuss the principal features of pairing in a very dilute
two-dimensional mixture of fermions and bosons, characterized by their 
masses $m_F$ and $m_B$; 
densities $\rho_F = k_F^2/4\pi$ and $\rho_B$; 
and chemical potentials $\mu_F \approx e_F = k_F^2/2m_F = 2\pi\rho_F/m_F$ 
and $\mu_B$. 
A fermion-boson mixture is an experimentally relevant situation, 
because at low density and temperature 
the most important contribution to the scattering amplitude is due to $s$-wave
collisions which, in the case of spin-polarized fermions,
are forbidden by Pauli's principle.
As a consequence, it is difficult to cool a sample of spin-polarized fermionic 
atoms to reach the temperatures needed to observe quantum degeneracy. 
This problem may be overcome by sympathetically cooling the fermions with 
a gas of bosons, so that the fermions cool down by interacting with the bosons
\cite{cool,tru01,mod02}.

We assume in the following the idealized zero temperature case.
The energy gaps characterizing the pairing 
can always be converted into critical temperatures by 
multiplying with the factor $\gamma/\pi \approx 0.567$ 
as in three dimensions~\cite{temp}.
We also assume $\mu_B \ll \mu_F$, which will be justified later.


Pairing in two dimensions has the peculiar feature that, for an attractive
$s$-wave interaction between two different fermionic species, a bound state
(of binding energy $E_b$) is always present 
and therefore the system enters the strong-coupling regime at sufficiently
low density \cite{schmitt,randl,rand,baran},
forming a Bose condensate of fermion pairs characterized by
\begin{subequations}
\bea
 \mu_F &\ra& e_F - E_b/2 \:,
\\
 \de_0 &\ra& \sqrt{2 E_b e_F} \:,
\label{e:sc}
\eea
\end{subequations}
where $\de_0$ is the $s$-wave pairing gap.
This strong pairing contrasts with the 3D 
weak-coupling behavior, 
where the gap vanishes exponentially with $k_F \rightarrow 0$. 
Moreover, the strong-coupling situation implies that 
the pairing gap is quite insensitive to
an asymmetry in the population of the two species \cite{nozieres}, 
contrary to the 3D case, 
where even a minute excess of particles of one species reduces considerably 
the gap size due to the effects of Pauli blocking in the gap
equation \cite{old,block}.

We therefore exclude in the following this ``trivial'' case,
and focus on the situation where $s$-wave pairing is not possible, 
either due to a repulsive $s$-wave interaction,
or when treating a system of identical 
(spin-polarized) fermions.
The next possibility concerns the $p$-wave pairing gap, 
$\de_1 \equiv \de_{L=1}(k_F)$,
which in the low-density limit 
is given by the weak-coupling result \cite{rand}
\be
 {\de_1 \over \mu_F} = 
 c_1 \exp\!\left[-{2\pi\over m_F T_F}\right] \:,
\label{e:wc}
\ee
where $c_1$ is a constant of order unity
and
\be 
 T_F = T_{k_F k_F}^{(L=1)}(2\mu_F)
 = \int_{0}^{\pi} {d\phi \over \pi} \cos{\phi} 
 \lla \kvp | T(2\mu_F) | \kv \rra
 \ ,\ |\kv|=|\kvp|=k_F
 \ ,\ \cos{\phi} = \bm{\widehat{k}}' \cdot \bm{\widehat{k}}
\label{e:tf}
\ee
is the relevant $T$-matrix element of the interaction.

\begin{figure}
\includegraphics[height=4.5cm]{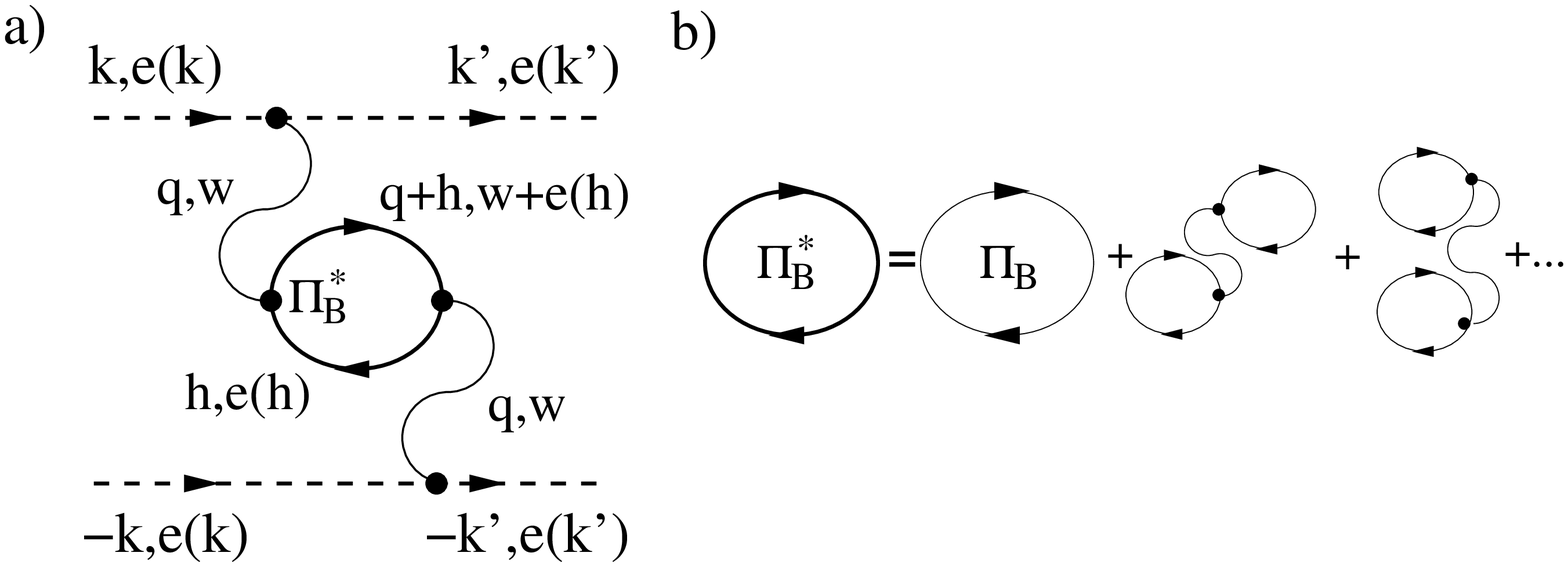}
\caption{
  (a) Polarization interaction $\Gamma$ between two fermions (dashed lines)
  mediated by the presence of bosons (solid lines). 
  The labels indicate the momentum and energy of each line.
  For condensate bosons and fermions
  on the Fermi surface, $\hv=\zv, \omega=0$. 
  (b) Diagrams contributing to the boson bubble in RPA; 
  the last one is an example of a backwardgoing diagram, negligible when
  $\mu_B\ra0$. 
  Here, thick solid lines are full propagators, thin solid lines are free
  propagators, and wiggles represent interactions.
}
\label{f:dia}
\end{figure}

We now analyze the pairing force mediated by the surrounding bosons.
Assuming for the moment that a direct fermion-fermion interaction is absent,
the relevant interaction to leading order in density,
to be used in Eq.~(\ref{e:wc}), is $T_F = \Gamma_F$, where $\Gamma_F$ 
is the boson-mediated irreducible polarization interaction,
schematically represented in Fig.~\ref{f:dia}.
Within the range of the weak-coupling formula, it is sufficient to 
consider in- and out-going fermions on the Fermi surface 
and on their energy shell, cf.~Eq.~(\ref{e:tf}).
Therefore, the energy transfer from one fermion to the other vanishes:
$\omega=0$.
For the time being we assume for simplicity boson-fermion ($BF$) and 
boson-boson ($BB$) $T$-matrices  
that can be considered constant in the low density limit, 
as in the three-dimensional case.
In two dimensions this is, however, not any more true \cite{adhi,morgan}, 
and the correct treatment will be discussed further below.

With this assumption, the relevant interaction kernel reads at 
low density \cite{viv}
\be
 \lla \kvp \left| \Gamma_{F} \right| \kv \rra = 
 \tf^2 \Pi^*_B(|\kv'-\kv|)
\label{e:wtot}
\ee
with the bosonic RPA propagator
\be
 \Pi^*_B(q) = { \Pi_B(q) \over 1 - \tb\Pi_B(q) }
\label{e:pis}
\ee
and the bosonic static Lindhard function
\be
 \Pi_B(q) = -{4m_B\rho_B\over q^2} \:.
\label{e:pi}
\ee
We have neglected the influence of the fermions on the properties of the Bose
condensate.
We remark at this point that due to the $1/q^2$ dependence of the 
two-dimensional Lindhard function, the RPA has necessarily 
to be performed in order to avoid divergencies.
The situation is similar to the electron gas where, however, 
the interaction is singular.

\begin{figure}
\includegraphics[totalheight=4.5cm]{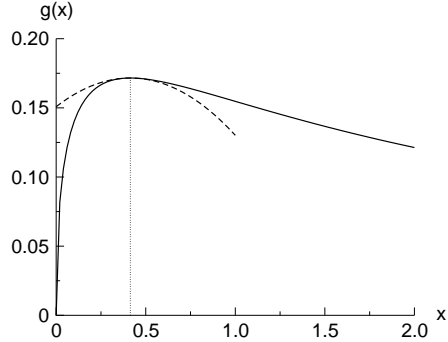}
\caption{
  The function $g$ appearing in Eq.~(\ref{e:g}) (solid line), together with
  the parabolic approximation (dashed line) around its maximum (indicated by 
  the dotted vertical line).}
\label{f:g}
\end{figure}

Projecting out the $L=1$ partial-wave $FF$ interaction, one obtains in
particular
\bea
 \Gamma^{(L=1)}_{k_F k_F} &=&
 {\tf^2}
 \int_{0}^{\pi} {d\phi \over \pi} \cos{\phi} \; 
 \Pi^*_B(q=\sqrt{2(1-\cos{\phi})}k_F)
\nonumber\\
 &=& -{\tf^2}\,  {x\over \tb}
 \int_{0}^{\pi} {d\phi \over \pi} { \cos{\phi} \over x + 1 - \cos{\phi}}
\nonumber\\
 &=& -{\tf^2 \over \tb} g\left( x \right) \:,\quad 
 x ={2m_B \tb \rho_B \over k_F^2} = {m_B \tb \over 2\pi} {\rho_B\over\rho_F}
\label{e:linkk}
\eea
with
\be
 g(x) = {1+x \over \sqrt{1+2/x}} - x \:.
\label{e:g}
\ee
This function is plotted in Fig.~\ref{f:g}.
It has a maximum located at
($x=\sqrt{2}-1 \approx 0.414, g = 3-2\sqrt{2}\approx 0.172$), 
and can in its vicinity be approximated by a parabola,
as shown by the dashed line in the figure.
This translates into a sharp Gaussian peak for the gap
function, according to Eq.~(\ref{e:wc}).

Therefore, when increasing the boson density for fixed $\rho_F$,
the induced fermionic attraction and thus also the pairing gap would 
reach a maximum for 
\be
 {\rho_B\over\rho_F}= {0.414\times 2\pi \over m_B T_{BB}} \:.
\label{e:xopt}
\ee

\begin{figure}
\includegraphics[totalheight=4.cm]{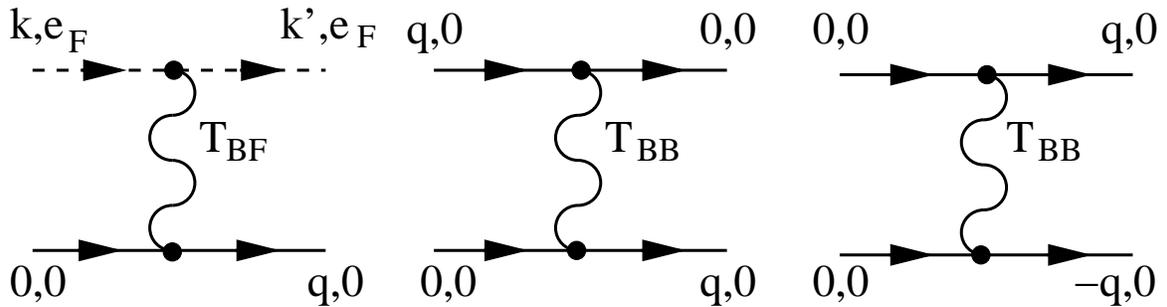}
\caption{
  Possible collision events in the mixture, according to Eq.~(\ref{e:col}). 
  Dashed lines denote fermions,
  solid lines bosons, and wiggles represent interactions. 
  The labels indicate the momentum and energy of each particle.}
\label{f:coll}
\end{figure}

However, in two dimensions the ($s$-wave) scattering matrices $\tf$ and $\tb$
cannot be considered constant, but vanish logarithmically with the
c.m.s.~energy $E$ of the two-particle state \cite{adhi,morgan}, i.e.,
\be
 \lla \kvp | T(\pv=\zv,E\ra 0) | \kv \rra \ra 
 {2\pi\over m_{\rm}} {1\over \ln{(E_0/|E|)}} \:,
\label{e:2dt}
\ee
where $m_{\rm}$ is the reduced mass of the colliding particles
and $E_0\gg E$ is a parameter (with dimensions of energy) characterizing 
low-energy scattering. 
Therefore, it is necessary to evaluate the c.m.s.~energy 
$E^2 = P_\mu P^\mu$ for the following situations
(sketched in Fig.~\ref{f:coll}):
\begin{subequations}
\bea
 \lla (\kvp,\mu_F)(\qv,0) | \tf | (\kv,\mu_F)(\zv,0) \rra &:& 
 E = {k_F^2\over 2m_F}{m_{BF}\over m_F} \:,
\label{e:col1}
\\
 \lla (\zv,0)(\qv,0) | \tb | (\qv,0)(\zv,0) \rra &:& 
 E = -{q^2\over 4m_{B}} \:,
\label{e:col2}
\\
 \lla (+\qv,0)(-\qv,0) | \tb | (\zv,0)(\zv,0) \rra &:& 
 E = 0 \:, 
\label{e:col3}
\eea
\label{e:col}
\end{subequations}
with $m_{BF}=m_B m_F/(m_B+m_F)$ the $BF$ reduced mass.
Therefore, within the approximation $\mu_B=0$ 
(or more precisely $\mu_B\ll \mu_F$),
only forwardgoing polarization diagrams 
(see Fig.~\ref{f:dia}b)
contribute to the induced 
$FF$ interaction.
This can be taken into account by replacing 
$\tb\Pi_B(q) \ra \tb(q)\Pi_B(q)/2$
in Eq.~(\ref{e:pis}), where now
\be
 \tb(q) = {4\pi\over m_B} {1\over \ln{(4 m_B E_{BB}/q^2)}} \: .
\label{e:tbb}
\ee
Also, the relevant boson-fermion interaction becomes
\be
 \tf(k_F) = {2\pi\over m_{BF}}{1\over \ln{(2m_F^2E_{BF}/m_{BF}k_F^2)}} \: .
\label{e:tbf}
\ee
Here $E_{BF}$ and $E_{BB}$ are the parameters characterizing
low-energy $s$-wave $BF$ and $BB$ scattering, respectively.

We obtain then
\bea
 \Gamma^{(L=1)}_{k_F k_F} &=&
 -{m_B \tf^2(k_F) \over 2\pi} h(x,y) \:,
\nonumber\\
 h(x,y) &=&
 \int_{0}^{\pi} {d\phi \over \pi} 
 { \cos{\phi} \over  (1 - \cos{\phi})/x - 1/\ln{[(1 - \cos{\phi})/y]} } \:,
\nonumber\\ &&
 \quad  x = {4\pi\rho_B \over k_F^2} = {\rho_B\over\rho_F}\:,
 \quad  y = {m_B E_{BB} \over m_F \mu_F} \:.
\label{e:h}
\eea
with the condition $y \gg 1$ for Eq.~(\ref{e:tbb}) to be valid.

\begin{figure}
\includegraphics[totalheight=9.cm,angle=0,bb=230 520 230 820]{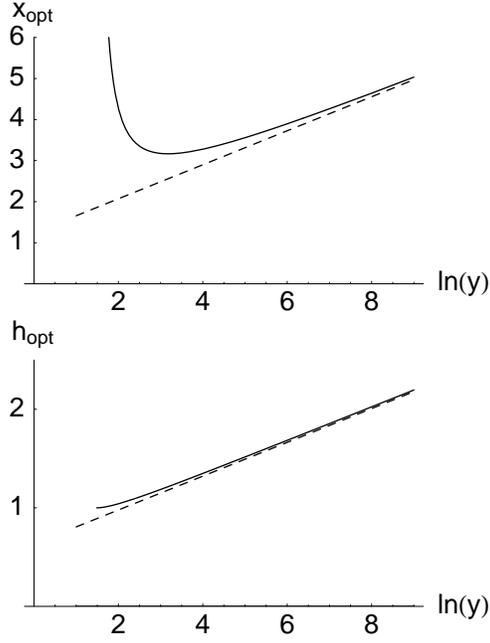}
\caption{
  The optimal values $x_{\rm opt}$ and $h_{\rm opt}$ 
  for the pairing interaction, Eq.~(\ref{e:h}).
  The dashed lines indicate the asymptotic behavior,
  Eq.~(\ref{e:opt}).}
\label{f:h}
\end{figure}

Varying the boson density (i.e., $x$) for a constant fermion density ($y$),
one observes again a maximum at a certain ratio $x_{\rm opt}(y)$.
The optimal ratio $x_{\rm opt}$ as well as the corresponding value 
$h_{\rm opt}$ are plotted in Fig.~\ref{f:h} as functions of $y$.
At sufficiently large $y$ one obtains a quasi linear dependence on $\ln y$:
\begin{subequations}
\bea
 x_{\rm opt}(y) &\ra& 0.414\, (3.0 + \ln{y}) \:,
\label{e:optx}
\\
 h_{\rm opt}(y) &\ra& 0.172\, (3.7 + \ln{y}) \:. 
\label{e:opth}
\eea
\label{e:opt}
\end{subequations}
We remark that in fact the optimal ratio $ x_{\rm opt}$ corresponds to the one 
for a constant $\tb$, Eq.~(\ref{e:xopt}), 
when making the replacement 
\be
 \tb \ra {\tb(q=0.317k_F)\over 2} \:.
\ee
Thus position and value of
the maximum depend logarithmically on the Fermi momentum.
Taking all these facts into account, 
the value of the pairing gap under optimal conditions becomes
\be
 \ln{\de_1\over c_1\mu_F} \ra 
 - {m_{BF}^2 \over m_B m_F} \;
 {[\ln{(m_FE_{BF}/m_{BF}\mu_F)}]^2 \over 0.172\left[3.7+\ln{y}\right]}
 \:.
\label{e:optgap}
\ee
The induced interaction, Eq.~(\ref{e:h}),  
is to be compared with the direct low-density 
$p$-wave fermion-fermion interaction \cite{rand},
\be
 T_{k_F k_F}^{(L=1)}(2\mu_F) \approx  {4\over m_F} { \mu_F \over E_1} 
 \sim \rho_F \:,
 \label{e:directFF}
\ee
where $E_1$ is the parameter characterizing 2D low-density $p$-wave scattering.
Therefore, at sufficiently low fermion density, the boson-mediated attraction,
Eqs.~(\ref{e:h},\ref{e:opt}), becomes dominant,
since it depends only logarithmically on the fermion density.
For the same reason, any fermionic polarization corrections have also
been neglected.

We analyze finally the assumption $\mu_B \ll \mu_F$ that was made beforehand.
The boson chemical potential is determined by \cite{mub,lee}
\be
 \mu_B = \rho_B \tb(E=\alpha\mu_B) =
 {4\pi\rho_B\over m_B} {1\over \ln{(E_{BB}/\alpha\mu_B)}} 
 \ll \mu_F =  {2\pi\rho_F\over m_F} \:,
\ee
where $\alpha$ is of order unity
\footnote{
In the current literature there is no agreement on the precise value
of the constant $\alpha$, see Refs.~\cite{mub,lee}. 
For our purpose, this is however not relevant.}.
Since the logarithm in the low-density domain is always large,
we have the sufficient condition 
\be
 x=\rho_B/\rho_F \lesssim m_B/m_F \:.
 \label{e:cond}
\ee

\begin{figure}
\includegraphics[totalheight=6.cm,angle=0,bb=170 640 170 820]{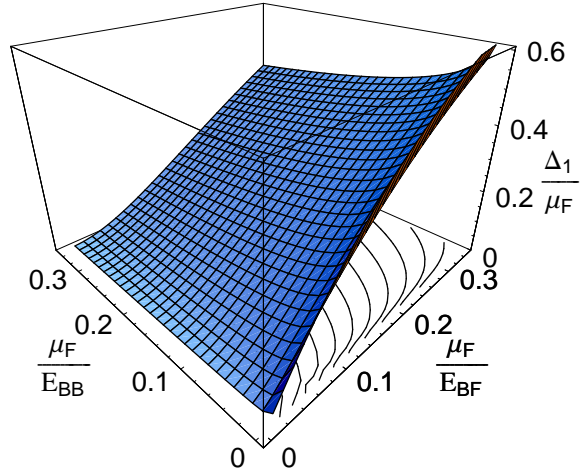}
\caption{
  The pairing gap for optimal boson concentration, Eq.~(\ref{e:optgap}),
  as a function of the fermion chemical potential.}
\label{f:d}
\end{figure}

In order to estimate typical sizes of the expected gap,
we plot in Fig.~\ref{f:d} the gap $\de_1/\mu_F$, 
according to Eq.~(\ref{e:optgap}),
as a function of the ratios
$\mu_F/E_{BB}$ and $\mu_F/E_{BF}$ 
(assuming for simplicity $c_1=1$ and $m_F=m_B$).
One notes that it is mainly the ratio $\mu_F/E_{BF}$ that determines the gap,
whereas the dependence on $\mu_F/E_{BB}$ is relatively weak.
Thus with fermion chemical potentials $\mu_F \lesssim E_{BF}$
quite large gaps $\de_1 \lesssim \mu_F$ could be achieved.

In order to translate this condition into experimental quantities,
we use the results of Refs.~\cite{petrov,lee}, 
relating the 2D scattering parameter $E_{BF}$ to the value of the 3D scattering
length $a_{BF}$,
for a boson-fermion system confined in a strongly anisotropic trap 
characterized by frequencies
$\om_\perp$ and $\om_z$
(here supposed to be the same for bosons and fermions), obtaining
\be
 {\mu_F \over E_{BF}} = 
 {\pi\over B} {\mu_F \over\om_z }
 \exp{\left(-\sqrt{2\pi}{\lz\over a_{BF}}\right)} 
 \:,
\ee
where $B\approx0.915$ and $\lz=1/\sqrt{2m_{BF}\om_z}$.
Since at the same time for a 2D situation the condition $\mu_F \ll \om_z$ 
must be fullfilled, one can only expect observable gaps
if the exponential term is not too small.
One can now distinguish two cases:
(i) $a_{BF}>0$: in this case the ratio $l_z/a_{BF}$ should be minimized 
as much as possible, i.e., for extremely strongly $z$-compressed traps, or for
systems with a very large BF scattering length (Feshbach resonance).
(ii) $a_{BF}<0$: in this case the exponential term is never small and
observable pairing can be expected
provided the ratio $\mu_F/\om_z$ is not too small.
Using the Thomas-Fermi approximation 
$\mu_F=\sqrt{2N_F}\omega_\perp$ for the chemical potential of a 
two-dimensional Fermi gas in a (in-plane) harmonic trap of frequency 
$\omega_\perp$, 
this last condition can be expressed by means of the fermion number 
and the trap asymmetry:
\be
 {\mu_F \over \om_z} = {\omega_\perp \over\omega_z } \sqrt{2N_F} \:.
\ee
Thus under favorable circumstances quite large $p$-wave pairing gaps 
of the order of the Fermi energy
seem to be achievable,
comparable to those of $s$-wave pairing in quasi-2D two-component Fermi
gases~\cite{baran}.
Unfortunately, more precise quantitative predictions cannot be made in 
this regime, since with $\mu_F\approx E_{BF}, E_{BB}$
also the asymptotic expression Eq.~(\ref{e:2dt}) becomes invalid.
It is worth noticing that the same effect in three dimensions is less
effective in increasing the size of the gap, and one expects
$\de_1/\mu_F\leq0.1$~\cite{hei00}.


Finally, we consider the problem of the phase-stability of 
boson-fermion mixtures,
which has been faced by different authors \cite{fel02,viv00,stab}
that reach similar conclusions. 
This problem has been studied for homogeneous as well as for trapped systems, 
but always in three dimensions, 
where the theoretical description is somehow easier than
in 2D because of the different behavior of the correponding $T$-matrices
at low energy. 
Here we briefly discuss the implications from the previous
studies \cite{fel02,viv00,stab}
that can be applied to our case, but a more precise
analysis would be of high interest.

According to Ref.~\cite{viv00}, in a boson-fermion mixture one can expect to
find one of three situations: 
(i) a fermionic phase and a bosonic phase,
(ii) a fermionic phase and a boson-fermion mixture, 
and (iii) a single uniform mixture. 
In case (i) there is no boson-fermion induced interaction and no 
sympathetic cooling. 
In case (ii) these problems are overcome, but only a
fraction of the fermions is efficiently cooled and can undergo the superfluid
transition. 
Therefore, the interesting situation is that of case (iii). 
This can be obtained if there is attraction between bosons and fermions 
(to avoid their spatial separation), 
but in this case the system may collapse due to
this same attraction \cite{fel02}. 
This will happen if, e.g., the number of
bosons exceeds some critical number $N_{\rm cr}$, which will depend on
$a_{BB}$ and $a_{BF}$. 
For a uniform system, we know that 
$a_{BB}>0$ is required in order to avoid the collapse of the boson component. 
This also stabilizes significantly the mixtures \cite{fel02}, 
even for $a_{BF}<0$.
As expected, the case $a_{BF}>0$ rapidly gives rise to spatial separation
of the two gases \cite{fel02}.

Applying these arguments to the mixtures used in typical experiments, 
we see that the case \Lin-\Lix\ 
(where the assumption $m_B=m_F$ is more adequate) 
with $a_{BB}=-1.5\;\rm nm$ and $a_{BF}=2.2\;\rm nm$ \cite{tru01}
does not correspond to the optimal stability conditions.
However, the presence of the trapping stabilizes the system so that
experiments can be performed. 
On the other hand, for the \Rb-\K\ mixture,
where $a_{BB}=5.2\;\rm nm$ \cite{mod02} and $a_{BF}=-2.2\;\rm nm$ \cite{kem02},
the stability conditions for the homogeneous case are fully satisfied. 


In conclusion, we have studied the characteristics of $p$-wave pairing in 
a two-dimensional boson-fermion mixture with repulsive (or absent) $FF$ 
$s$-wave interaction.
The boson-induced attraction between two fermions dominates at low density 
an eventual direct $FF$ $p$-wave force.
The induced pairing gap becomes maximal at a certain optimal boson-fermion
ratio.
In contrast to the three-dimensional case, this ratio itself increases when
decreasing the fermion density, due to the logarithmic energy dependence of
the $BB$ $T$-matrix at low density.
Using this optimal condition, we have estimated the size of the gap 
and find experimentally achievable values, in particular for systems
with a negative boson-fermion scattering length
such as the \Rb-\K\ mixture.


\medskip
We would like to acknowledge valuable discussions with M. Guilleumas
and M.A. Baranov.
This work was supported in part by the program DGICYT (Spain)
No.~BFM2002-01868.
J. M.-P. acknowledges support from a fellowship of the Generalitat
de Catalunya.


\newpage

\end{document}